\documentclass[10pt,twocolumn]{article}

\usepackage[utf8]{inputenc}
\usepackage{amsmath,amssymb}
\usepackage{graphicx}
\usepackage{cite}
\usepackage{authblk}
\usepackage{geometry}
\usepackage{abstract}  
\usepackage{titlesec}
\usepackage{caption}
\usepackage{hyperref}
\usepackage{float}

\usepackage{xcolor}

\usepackage{wasysym}

\usepackage{siunitx}

\usepackage{comment}

\titleformat{\section}{\large\bfseries}{\thesection.}{0.5em}{}
\titleformat{\subsection}{\normalsize\bfseries}{\thesubsection.}{0.5em}{}

\title{\textbf{Universal logic circuit for gate-controlled
superconductor-based switches operating at liquid-helium temperatures}}
\author[1,2]{Martin Berke}
\author[1,2]{Lőrinc Kupás}
\author[1,2,3]{Tosson Elalaily}
\author[1,2,4]{Máté Sütő}
\author[1,2,4]{Szabolcs Csonka}
\author[1,5]{Péter Makk}
\affil[1]{\small Department of Physics, Institute of Physics, Budapest University of
Technology and Economics, Műegyetem rkp. 3., H-1111 Budapest, Hungary}
\affil[2]{\small MTA-BME Superconducting Nanoelectronics Momentum Research Group, Műegyetem rkp. 3., H-1111 Budapest, Hungary}
\affil[3]{\small Low-Temperature Laboratory, Department of Applied Physics,
Aalto University School of Science, P.O. Box 15100, FI-00076, Aalto, Finland}
\affil[4]{\small Institute of Technical Physics and Materials Science, HUN-REN Centre for Energy Research, Konkoly Thege Miklós út 29-33., H-1121 Budapest, Hungary}
\affil[5]{\small MTA-BME Correlated van der Waals Structures Momentum Research Group, Műegyetem rkp. 3., H-1111 Budapest, Hungary}

\date{\today}

\begin{document}

\twocolumn[
  \maketitle
  \begin{onecolabstract}
    The observation of the gate-controlled supercurrent (GCS) effect in superconducting nanostructures initiated major research efforts toward the realisation of superconducting-based computing architectures. 
    Here we introduce a universal logic circuit that can be a promising superconducting building block of classical hybrid supercomputers. We demonstrate a functionally complete set of logic gates by realising the AND, OR, NOT and COPY gates. The general layout and scalability of our device, combined with recent experiments demonstrating fast switching and small voltage signals, make it a functional candidate in superconducting electronics. Our device enables the realization of all classical logic gates and the half-adder combinational logic circuit using at most three nanowires, each uniquely configured with two side-gate electrodes.
  \end{onecolabstract}
  \vspace{0.5cm}
]

\section{Introduction}
Since Gordon Moore’s 1965 prediction that the number of transistors on an integrated circuit doubles roughly every two years--known as Moore’s law--this trend has driven decades of progress in computing performance. However, in recent years, the scaling of CMOS-based devices has slowed due to fundamental physical limits, increased power consumption, and increasing cooling demands in modern supercomputers \cite{moore1998cramming,moore2003no,horowitz20141}.
The concept of a superconductor–semiconductor hybrid supercomputer offers a promising solution to these challenges, as elements operating in the superconducting state are inherently dissipationless \cite{holmes2013energy,soloviev2017beyond}. In this architecture, data processing is carried out using superconducting elements, while data storage relies on conventional semiconductor devices. However, a major obstacle for realizing such a hybrid platform is that most of the proposed superconducting devices are primarily controlled magnetically (via generated fluxes). This approach is problematic because it requires complex global bias networks with continuous static power dissipation, thereby severely limiting scalability and overall energy efficiency.

In 2018, De Simoni et al.\,\cite{de2018metallic} observed the monotonic suppression of the supercurrent in fully metallic (Ti) nanowires (NWs) by increasing the voltage on a closely placed gate electrode, and by further increasing the gate voltage, the device is completely switched to the normal (N) state. 
The marked effect of the gate voltage on the maximal supercurrent is referred as the gate-controlled supercurrent (GCS) effect \cite{ruf2024gate}.
The origin of the GCS effect in earlier works was attributed purely to the electric field \cite{de2018metallic,amoretti2022destroying,bours2020unveiling,chakraborty2023microscopic,chirolli2021impact,mercaldo2020electrically,solinas2021sauter,rocci2020gate,puglia2020electrostatic,paolucci2019magnetotransport,mercaldo2021spectroscopic}, but more recent works show that the effect is likely related to leakage current in the substrate at relatively high gate voltages \cite{ritter2021superconducting, elalaily2021gate, elalaily2023signature, elalaily2024switching, ritter2022out, basset2021gate, li2025incoherent, ruf2023effects}. Recently switching due displacement currents was demonstrated as well \cite{scherubl2025multimode}.

The effect of gate voltage on the superconducting (SC) NWs provides a superconducting voltage-controlled switch, the superconducting equivalent of a semiconductor transistor, with low power consumption and high switching speed, that was recently demonstrated \cite{scherubl2025multimode,buccheri2025microwave}.
The simple structure and voltage-controlled operation of these superconducting nanodevices can be scaled up with greater compatibility for interfacing with CMOS transistors in comparison to alternative superconducting devices \cite{buck1956cryotron, likharev1991rsfq, matisoo1966subnanosecond, mccaughan2014superconducting}.

Motivated by the promising advantages of the GCS effect, this work demonstrates a device architecture for logic circuits of minimal structural complexity governed by this phenomenon. The elementary building block consists of a superconducting NW accompanied with two gate electrodes. Building on this element, the proposed layout with three such building blocks (as shown in Fig.~\ref{fig:fig1}a)) is capable of implementing the complete set of classical logic functions and key combinational circuits, including AND, OR, NOT, COPY, NOR, XOR, NAND, as well as a half adder and a ring oscillator. 

Whereas, a conventional two-input CMOS NAND gate requires four transistors \cite{razavi2021fundamentals}, and other exemplary SC devices require a considerable number of building blocks to realise more complex circuits, such as the half-adder \cite{mccaughan2014superconducting}, the same functionality in our approach is implemented using only three superconducting NWs, highlighting the architectural efficiency of the proposed platform.

In this work, therefore we present the operating principle, fabrication, and initial experimental characterization of simple logic gates realized using this universal superconducting circuit layout. Our work demonstrates the usability of GCS effect for logical circuits and opens up the field to more complex device architectures operating at liquid-helium temperatures.

\section{Device outline}
Devices were fabricated using standard nanofabrication techniques. Undoped Si substrates with a $\SI{290}{nm}$ thick $\text{SiO}_2$ layer were patterned by electron-beam lithography to define NWs, side gates, and contacts. A 30 nm NbTiN film was then deposited by reactive DC magnetron sputtering, followed by lift-off; more details on the fabrication procedure is given in the Supplementary Material.
The NWs are made of NbTiN, which offers a relatively high critical temperature, ensuring that the device remains fully operational at liquid-helium temperatures \cite{thoen2016superconducting}. 
Furthermore, by utilizing long nanowires, our samples exhibit a relatively high normal-state resistance, a direct consequence of the intrinsically high resistivity characteristic of disordered transition-metal nitride films \cite{myoren2001properties}. Recent works also discuss the possibility of using highly resistive niobium Dayem bridges to obtain substantial output voltage \cite{ruf2024high}.

\begin{figure}[]
\centering
\includegraphics[width=0.8\columnwidth]{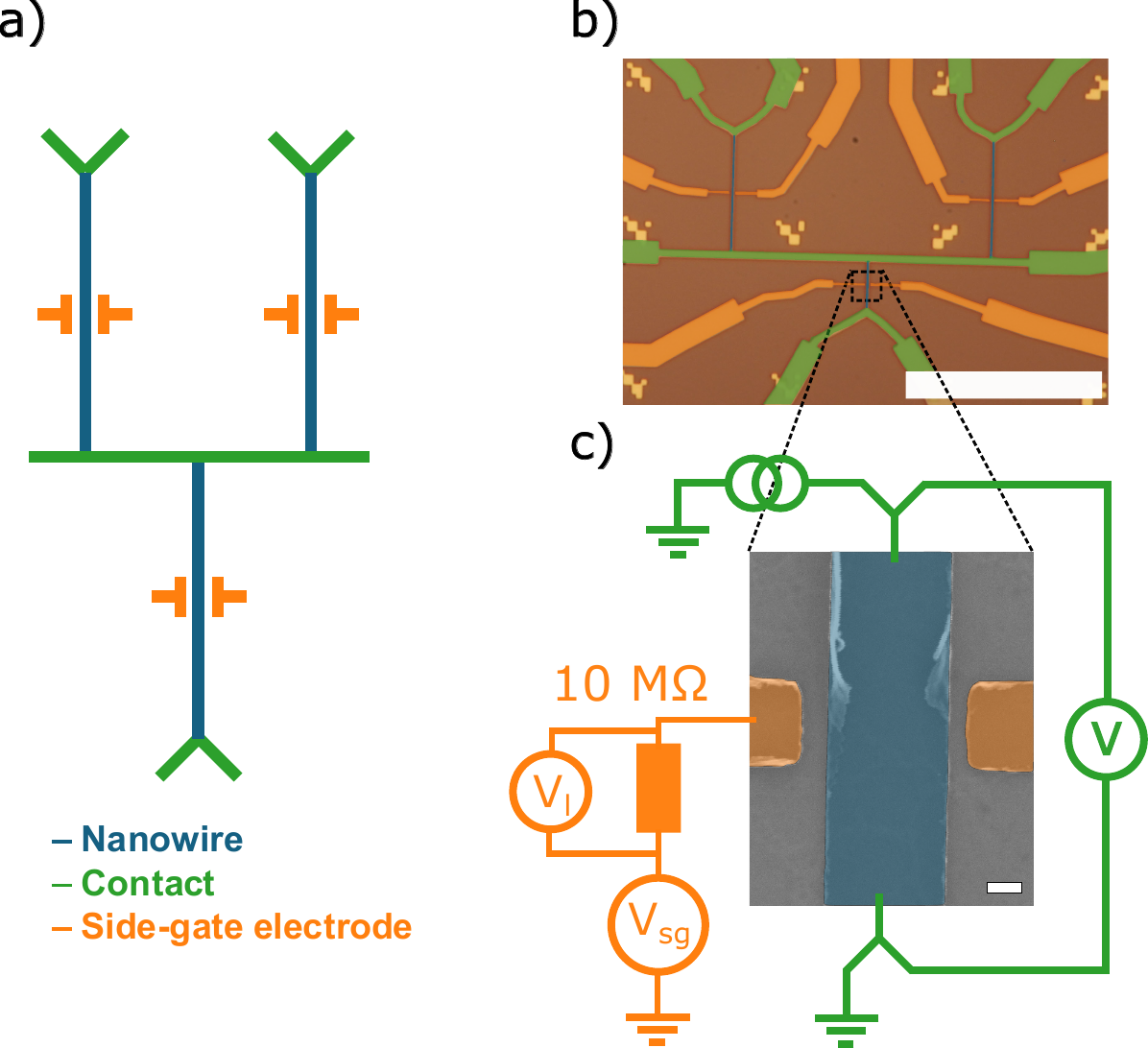}
\captionsetup{labelformat=empty}
\caption{Fig. 1. \textbf{Device configuration} (a) The device consists of three NWs (blue) that are connected to a common line (green) at one end, and there are two more contacts (green) at the other end of each NW. Each NW is controlled by two side gate electrodes (orange). (b) Optical microscope image of a device. Scale bar: \SI{25}{\micro\meter} (c) SEM image of the active region of the device showing the wire and the gate electrodes Scale bar: \SI{100}{\nano\meter}.}
\label{fig:fig1}
\end{figure}

The schematic of the fabricated device geometry is shown in Fig.~\ref{fig:fig1}a). Each device consists of three NWs (blue) that are connected to a common line (horizontal green line) at one end, and there are two more contacts (green) at the other end of each NW. 
Additionally, there are two side gate electrodes (orange) for each NW on which a voltage signal can be applied. 
The false coloured optical microscope image of a fabricated device is shown in Fig.~\ref{fig:fig1}b) wit the same colour coding, and the gated region with a schematic representation of the measurement setup of a NW is shown in the scanning electron microscope (SEM) image in Fig.~\ref{fig:fig1}c). The typical separation of the NW and gate electrodes is \SIrange{50}{100}{\nano\meter}.

\section{Experimental results}
We first discuss the basic characterisation of the device, then explain the logic operation principle, and finally demonstrate a functionally complete set of basic logic gates.

Measurements were carried out at \SI{4}{\kelvin}, the critical temperature of device is $T_c\approx$ \SI{12}{\kelvin}. For the measurements presented in the following sections, the sample was directly immersed in the liquid He.
The bias current was applied via a \SI{200}{\kilo\ohm} ohmic resistor in series with the NWs, and the leakage from the gate electrode was probed by measuring the voltage drop $V_{\text l}$ on a \SI{10}{\mega\ohm} ohmic pre-resistor in series with the gate electrode, as indicated in Fig.~\ref{fig:fig1}c).
We used only two out of the total of three NWs in our measurements. First we present the $V - I$ curve and the quantitative analysis of gating measurements for one of these wires (NW1), as the qualitative behaviour of the other wire (NW2) is the same. 
A more detailed analysis, including NW2 and the sample charachterization is given in the Supplementary Material.

\begin{figure}[H]
\centering
\includegraphics[width=1.\columnwidth]{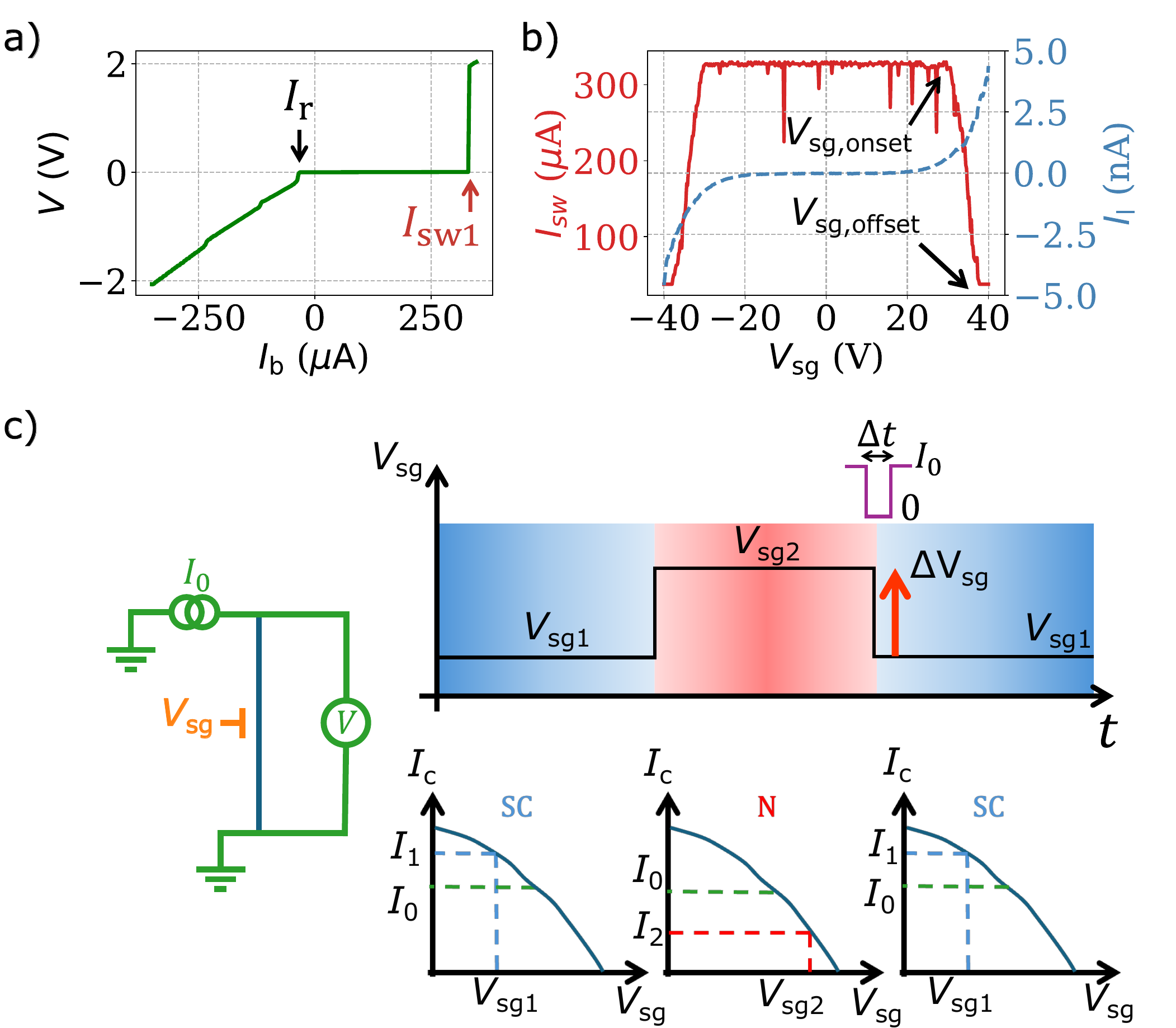}
\captionsetup{labelformat=empty}
\caption{Fig. 2. \textbf{Suppression of the supercurrent with gating the device and logic operation principle} (a) $V-I$ curve of a NW measured at \SI{4}{\kelvin}. (b) The critical current of a NW as a function of the applied gate voltage on one of the side-gate electrodes (red) and the leakage current calculated from the voltage drop measured across a pre-resistor (blue). With increasing gate voltage on the side-gate electrode, the critical current of the NW can be suppressed, accompanied by a large enhancement in the leakage current. (c) An external voltage signal can switch a NW between the SC and N states.  For devices where the onset and offset of the gating effect correspond to gate voltages largar than $\SI{5}{\volt}$, the logical value is carried by $\Delta V_{\text{sg}} = V_{\text{sg2}}-V_{\text{sg1}}$.}
\label{fig:fig2}
\end{figure}

\subsection*{A. Gate influence}
We first investigated the GCS effect in our device by measuring the $V - I$ characteristics as a function of the applied gate voltage on the side-gate electrode $V_{\text{sg}}$ for both polarities in a quasi-four-probe setup for a NW, as shown in Fig.~\ref{fig:fig1}c).
The effect of the applied gate voltage on the switching current is similar to earlier observations \cite{amoretti2022destroying,bours2020unveiling,chakraborty2023microscopic,chirolli2021impact,mercaldo2020electrically,solinas2021sauter,rocci2020gate,puglia2020electrostatic,paolucci2019magnetotransport,mercaldo2021spectroscopic, de2018metallic,ruf2024gate,ritter2021superconducting, elalaily2021gate, elalaily2023signature, elalaily2024switching, ritter2022out, basset2021gate, li2025incoherent,scherubl2025multimode,ruf2023effects, koch2024gate}. 
In Fig.~\ref{fig:fig2}a) we show the $V-I$ curve at $V_{\text{sg}} = 0$, the switching current is approximately $I_{\text{sw1}} \approx \SI{332}{\micro\ampere}$ and the resistance in the dissipative state is $R_\text{n} \approx \SI{5.86}{\kilo\ohm}$. 
The device has a rather large hysteresis; the retrapping current is $I_{\text{r}} \approx \SI{35}{\micro\ampere}$ (current values are indicated by arrows). 
This value is an order of magnitude smaller than the switching current, indicating the presence of strong Joule-heating, which is compatible with the large normal state resistance and switching current values.
The voltage drop in the N state is much higher compared to many other devices investigated so far \cite{alegria2021high,basset2021gate,amoretti2022destroying,bours2020unveiling,de2018metallic,de2019josephson,elalaily2021gate,elalaily2023signature,elalaily2024switching,golokolenov2021origin,chirolli2021impact,mercaldo2020electrically,mercaldo2021spectroscopic,paolucci2019magnetotransport,solinas2021sauter,rocci2020gate,ritter2022out,ritter2021superconducting,puglia2020electrostatic}, making these devices promising for applications. 
The voltage drop in the N state exceeds noise floor of cryogenic wiring and electronics eliminating the need for local amplification and enables simpler interfacing with existing CMOS technology.
The output voltage of our NbTiN NW which is $\approx\SI{2}{\volt}$ for bias current values slightly above the switching current, whereas in other superconducting devices the corresponding voltage is typically limited to the millivolt range.

The influence of the applied gate voltage on the side-gate electrode is shown in Fig.~\ref{fig:fig2}b). The switching current begins to be suppressed from $V_{\text{sg,onset}} \approx \SI{30}{\volt}$ and diminishes at $V_{\text{sg,offset}} \approx \SI{40}{\volt}$ (indicated by arrows). 
In the gating measurements, we also measured the leakage current between the side-gate electrode and the ground by recording the voltage on a pre-resistor in series with the gate electrode, as it is shown in Fig.~\ref{fig:fig1}c).
At gate voltages, where the switching current is suppressed, a corresponding enhancement in the leakage current is observed, as shown in Fig.~\ref{fig:fig2}b) (blue curve), similarly to other works \cite{basset2021gate,elalaily2021gate,elalaily2023signature,elalaily2024switching,ritter2022out,ritter2021superconducting}.
Since the temperature is only approximately 1/3 of the critical temperature, thermal fluctuations can be observed \cite{bezryadin2013superconductivity} and explain small scattering on the $I_{{\text{sw}}}-V_{{\text{sg}}}$ plateau, whereas imperfect electrical isolation of the setup could also play a role.

\subsection*{B. Logic operation principle}
To perform logic operations, we first identify the setpoint for operating the device for which the setup shown in Fig.~\ref{fig:fig2}c) is considered. 
The device is operating at a working point where the side-gate voltage is $V_{\text{sg}}=V_{\text{sg1}}$ and the corresponding critical current is $I_{\text{c1}}$. The sample is current biased with bias current $I_0$ satisfying $I_0 < I_{\text{c1}}$, as shown in the first part of the time series in Fig.~\ref{fig:fig2}c). 
In this regime, the NW will be superconducting, with zero output voltage (indicated by the blue background). 

By increasing the gate voltage to $V_{\text{sg}}=V_{\text{sg2}}$, for which the corresponding critical current $I_{\text{c2}}$ satisfies $I_{\text{c2}}<I_0$, the device switches to the normal state with finite output voltage, as shown with red background in the middle part of the time series in Fig.~\ref{fig:fig2}c).

Finally, if the gate voltage is reduced to $V_{\text{sg}}=V_{\text{sg1}}$, the device can be switched back to the superconducting state as it is illustrated in the last part of the schematic time series in Fig.~\ref{fig:fig2}c).
If the device exhibits finite hysteresis due to Joule heating, then during the latest switching event, the bias current must be briefly reduced below the retrapping current for a short time interval $\Delta t$ (see Fig.~\ref{fig:fig2}c). For devices with minimal hysteresis in the $V - I$ characteristic, this condition is not restrictive.
The reset during the operation is also required for other exemplary SC based devices \cite{mccaughan2014superconducting}.

It is clear that with an applied gate voltage, the device can be switched between two states, which can be distinguished via voltage measurement; thus, our device consists of superconducting voltage-controlled switches. 
For the realization of certain logic gates, both gate electrodes of a NW are used. Since the sum rule of the gate voltages is unconventional \cite{puglia2021gate, elalaily2023signature}, the operating point must be chosen carefully.
More information on the chosen working points for the logic gates can be found in the Supplementary Material.

\subsection*{C. Logic gate operation}
To demonstrate the suitability of our device for digital applications, we used it to realise AND/OR/NOT and COPY gates. 
The operating principle for the rest of the logic gates is shown in the Supplementary Material.
All four logic gates require different connections of the nodes of the universal circuit scheme introduced previously.
For the measurements presented in this subsection, we used only two NWs; therefore we only include the active NWs and side-gate electrodes in the representative figures. 
The critical current, the retrapping current and the normal state resistance of the second wire is
$\mbox{$I_{\text{sw2}} \approx \SI{252}{\micro\ampere}$}$,
$\mbox{$I_{\text{r2}} \approx \SI{35}{\micro\ampere}$}$ and
$\mbox{$R_\text{n} \approx \SI{3.69}{\kilo\ohm}$}$.
The inputA and inputB signals in the upcoming discussion represent logic signals.
The details on the choices of the working points and more information on the NWs can be found in the Supplementary Material.

\begin{figure*}[t] 
    \centering
    \includegraphics[width=\textwidth]{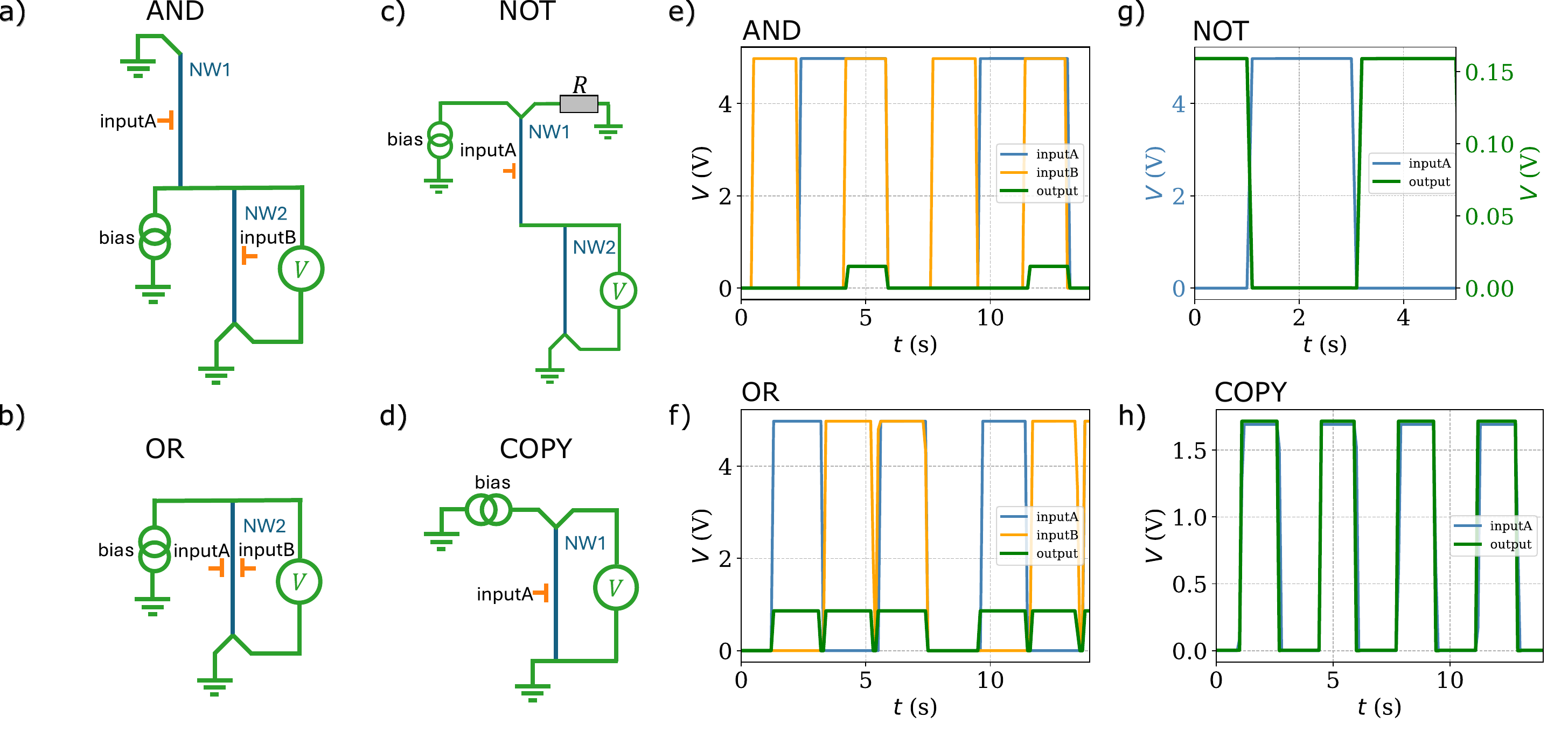}
    \captionsetup{labelformat=empty}
    \caption{Fig. 3. \textbf{Logic gates operation for the AND, OR, NOT and COPY gates.} 
    (a)-(d) The schematic circuit diagram for the AND, OR, NOT and COPY gates. The input signals inputA and inputB represent the \textit{additional} gate voltage on top of the gate voltage setting the working point on panels (e)-(h).
    (e) The logic operation of the AND gate. There is a clear increase in the output voltage (green) as the function of time, when both input signals  (blue, orange) are high.
    (f) The logic operation of the OR gate. The output voltage (green) as the input signals (blue, orange) are changed with respect to time. A clear increase in the output voltage can be seen, when any or both input signals are high.
    (g) The logic operation of the NOT gate. The output voltage (green) as the input signal (blue) is changed with respect to time. The output voltage is high when the input signal is low.
    (h) The logic operation of the COPY gate. The output voltage (green) as the input signal (blue) is changed with respect to time. By choosing an appropriate set point for operation, not just the logical value of the signal, but the value itself could be reproduced.
}
    \label{fig:fig3}
\end{figure*}

For the realisation of the AND gate, two NWs and one side-gate electrode for each NW are needed, as shown in Fig.~\ref{fig:fig3}a). 
The bias current is applied to the common line, whereas the other end the the NWs are grounded, and the two inputs are applied to the two side-gate electrodes. Finally, the output is measured on the common line with respect to the ground. 
If the inputs are such that both NWs are switched to the normal state, the output will be finite. 
In all other cases when one or both of the inputs is such that any of the wires is in the SC state, the bias current will flow to the ground through a superconducting element and the voltage drop will be zero. 

In Fig.~\ref{fig:fig3}e) we demonstrate the operation of the AND gate, where the inputs on the gate electrodes are shown with blue and orange colours, and the output is plotted with green as a function of time. 
The measured output is low if either inputA or inputB are low, and it is high if both of the input signals are high, therefore it is clear that the device is capable of reproducing the truth table of the AND gate.

For the realisation of the OR gate only one NW and both of the sidegate electrodes are required, as shown in Fig.~\ref{fig:fig3}b).
One end of the wire is connected to the ground, while the bias current is applied on the other end.
The inputs are applied to the two side-gate electrodes, and the output is measured across the wire. 
If the applied gate voltage on any of the side gate electrodes switches the device to the normal state, then the output voltage will be finite.
In the case where the applied gate voltages on the gates do not switch the device, the output voltage will be zero.

In Fig.~\ref{fig:fig3}f) we plot the inputA and inputB with colours blue and orange respectively, and we plot the measured output voltage with green as a function of time. 
The measured output is high if either inputA or inputB is high, and it is low if both of the input signals are low, therefore device is capable of reproducing the truth table of the OR gate.

Fig.~\ref{fig:fig3}c) shows the possible realization of the NOT gate, for this logic gate, an extra ohmic resistor connected parallel with the NWs is required, the specific value in our experiment is \SI{1}{\kilo\ohm}. 
As shown in Fig.~\ref{fig:fig3}c), the output is measured on the second NW. 
By applying the appropriate $I_0$ bias current and input, an operation point can be achieved where the first wire is in the SC state and the second wire is in the N state. 
This can happen if the current $I$ in the NWs is larger than the critical current of the second NW, i.e. $I>I_{\text{c2}}$ and the current $I$ also satisfies $I<I_{\text{c1}}$.
The extra ohmic resistor controls the current value $I$.
When the input logic signal causes the first nanowire to switch to the normal state, its normal-state resistance becomes larger than the resistor value, i.e.,
$R_{\mathrm{n1}} > R$.
As a result, more current is diverted into the resistor branch. This reduces the current flowing through the branch containing the second nanowire below its critical current, allowing the second nanowire to switch to the superconducting state.
For small input amplitudes, the output remains finite. However, a sufficiently large input switches the first wire to the N state, decreasing the current in that branch and inducing a transition of the second wire to the SC state. Therefore, the operation fulfills the truth table of a classical NOT gate.

The measurement is shown in Fig.~\ref{fig:fig3}g), the input signal is plotted with blue, and the output signal is shown in green as a function of time in the lower panel clearly reproducing the truth table of the NOT gate.

Finally, we present the principle of the operation of the COPY gate in Fig.~\ref{fig:fig3}d). If the input on the side-gate electrode is small, the measured voltage across the NW will be zero; however, once the NW is switched to the normal state, the output voltage will be finite. One can copy analogue values of the applied gate voltage, not just the logical value, as the normal state resistance and the large critical current of the devices allow for output voltages on the order of 1 V (for more information see Supplementray Material).

The measurement is shown in Fig.~\ref{fig:fig3}h), the input signal is plotted with blue, and the output signal is shown in green as a function of time.
The measured output signal follows the logical information carried by the input signal; thus, showing the COPY gate operation clearly.

Further information about other logic gates and circuits that can be realized with this device archticture is given in the Supplementary Material.

\section{Conclusion}
To summarise, we have described a simple fabrication process of a universal electronic circuit, consisting of three NWs and two side-gate electrodes, which is capable of performing classical logic operations and combinational operations by choosing the appropriate connections on a connector box. Our device is compatible with liquid helium temperatures and a large normal-state voltage in the range of a few volts can be measured upon switching to the normal state, desired for the integration within CMOS technology.
We showed that the critical current of a NW can be controlled by an applied gate voltage on a nearby side-gate electrode; thus, the device operates as a voltage-controlled superconducting switch.
As a next step, we have shown a functionally complete set of basic logic gates by investigating configurations which are capable of reproducing the truth tables of the AND, OR, NOT and COPY logic gates. 

The universality of the device has an important role when one wants to realise logic gates or combinational circuits, which with standard CMOS technology and other SC devices would require many devices. 
Our layout allows the realisation of all classical logic gates and the half-adder combinational circuit with three NWs and two side gate electrodes for each NW.
It is also important that one can easily switch between different logic gates by choosing the appropriate connections.

The superconducting state allows dissipationless operation, whereas the control via gate electrodes suggests a scalable architecture.
The structure of our nanodevices can be scaled up with greater compatibility for interfacing with CMOS transistors compared to other superconducting devices, and together with recent measurements confirming high-speed performance, this lays the groundwork for potential applications of the GCS effect \cite{puglia2021gate, ruf2024gate,ritter2021superconducting, scherubl2025multimode}. 
Finally, gate-controlled superconducting architectures can also play an important role in quantum computing, where the increasing number of qubits and control lines calls for on-chip logic circuits and multiplexing \cite{paghi2025supercurrent}. 

\section*{Data availability}
The data in this publication are available in numerical form at: \url{https://doi.org/10.5281/zenodo.21066259}.

\section*{Acknowledgments}
This work was funded by the EU’s Horizon 2020 research and innovation program under grant SuperGate
network (964398), the EIC Pathfinder Challenge grant
QuKiT (101115315), by the European Research Council ERC project Twistrain, the COST Action CA21144 (SUPERQUMAP). 
This work was supported by Jane and Aatos Erkko foundations as part of the SuperC collaboration, QMAT RCF CoE grant no. 374170.
This research was supported by the Ministry of Culture and Innovation and the National Research, Development and Innovation Office within the Quantum Information National Laboratory of Hungary
(Grant No. 2022-2.1.1-NL-2022-00004).  M. B. was also partially supported by the Doctoral Excellence
Fellowship Programme (DCEP), funded by the National Research Development and Innovation Fund
of the Ministry of Culture and Innovation and the Budapest University of Technology and Economics,
under a grant agreement with the National Research, Development and Innovation Office.

\section*{Competing Interests}
The authors declare that a patent application related to the content of this publication has been filed in Hungary.

\section*{Author contribution}
M. B., M. S. and T. E. fabricated the device. M. B., L. K. and T. E. performed the measurements. Data analysis was performed by M. B. and L. K.. The project was guided by P. M. and S. C.. The manuscript was written by M.B. with input
from all authors.

\bibliographystyle{ieeetr}
\bibliography{references}

\clearpage
\onecolumn

\setcounter{figure}{0}
\begin{titlepage}

    \centering
    
    \vspace*{2cm}
    
    {\LARGE\bfseries Supplementary Information\par}
    
    \vspace{1cm}
    
    {\large for the article\par}
    
    \vspace{0.4cm}
    
    {\Large\itshape Universal logic circuit for gate-controlled
superconductor-based switches operating at liquid-helium temperatures\par}
    
    \vspace{1.2cm}
    
    {\large

    Martin Berke$^{1,2}$, Lőrinc Kupás$^{1,2}$, Tosson Elalaily$^{1,2,3}$, Máté Sütő$^{1,2,4}$, Szabolcs Csonka$^{1,2,4}$, Péter Makk$^{1,5}$\par}
    
    \vspace{0.8cm}
    
    {\small

    $^{1}$ Department of Physics, Institute of Physics, Budapest University of
Technology and Economics, Műegyetem rkp. 3., H-1111 Budapest, Hungary\par
    $^{2}$MTA-BME Superconducting Nanoelectronics Momentum Research Group, Műegyetem rkp. 3., H-1111 Budapest, Hungary\par
    $^{3}$Low-Temperature Laboratory, Department of Applied Physics,
Aalto University School of Science, P.O. Box 15100, FI-00076, Aalto, Finland\par
    $^{4}$Institute of Technical Physics and Materials Science, HUN-REN Centre for Energy Research, Konkoly Thege Miklós út 29-33., H-1121 Budapest, Hungary\par
    $^{5}$MTA-BME Correlated van der Waals Structures Momentum Research Group, Műegyetem rkp. 3., H-1111 Budapest, Hungary\par
    }
    
    \vfill

\end{titlepage}

\clearpage

\section*{\large Sample fabrication}
Our devices were fabricated via a single-step electron beam lithography (EBL) and a reactive, DC magnetron sputtering step, finished by a lift-off process. 
\begin{itemize}
    \item The undoped Si chips with an oxide layer 290/285 nm thickness were covered with positive e-beam resist (PMMA, AR-P 669, 600K, 2.5 \% diluted), baked at 170-185 $^{\circ}$C for 3-5 minutes.
    \item Electron beam exposure in a Raith150 system, 20 kV, 840 $\mu$C/cm$^2$ dosage uniformly, 10 $\mu$m aperture for fine structures and  30 $\mu$m aperture for the contacts.
    \item Development in cold MIBK:IPA 1:3 for 60 s, rinsed in IPA for 30 s.
    \item Metallization: Gas flow rates: Ar: 10 sccm, N: 2.4 sccm. Pressure: 2.7 mTorr. Power, time: 150 W, $\approx$ 6 min. Bias plasma power: 5 W. 
\end{itemize}

\section*{\large Measurement setup}

Measurements were performed in liquid helium to maintain cryogenic temperatures.
We used a Basel Precision Instruments SP927 DAC as a voltage source together with Basel HV-Amplifiers to apply the gate voltages. The bias current was applied via a 200\,k$\Omega$ ohmic resistor in series, and finally, we measured the leakage by measuring the voltage drop on a 10\,M$\Omega$ ohmic resistor in series with the gate electrode.
Voltages were measured via Keithley 2000/2001 multimeters.

\section*{\large Data processing}

To obtain the $I_{\mathrm{sw}}$--$V_{\mathrm{sg}}$ characteristics presented in the manuscript, a voltage threshold of 0.05~V was applied to determine the approximate switching current.

For the leakage current measurements, the offset at $V_{\mathrm{sg}} = 0$ was subtracted from the dataset, and the measured voltage signal was converted to current by dividing by the value of the series pre-resistor.

In the logic gate measurements, the Input~A and Input~B signals were shifted to zero for clarity in the graphical representation; no additional data processing was performed.

\section*{\large Number of samples tested}
More than five devices of this type were systematically characterized during the initial measurement phase. Owing to the setup limitations, the extended charactzerization and training required for logic gate operation was conducted on a reduced subset of devices (n = 2).

\section*{\large Sample characterization}

\subsection*{Critical temperature}
We estimated the critical temperature of the NbTiN thin film by monitoring voltage switching events in a weakly biased nanowire while varying the sample position relative to the He level. From two consecutive up–down sweeps, the average critical temperature was $T_{\mathrm{c}}=12.1,\mathrm{K}$; the measured values are listed in Table~\ref{tab:tempdata}.

\begin{table}[h!]
\centering
\begin{tabular}{|c|c|c|c|c|}
\hline
$T_1$ & $T_2$ & $T_3$ & $T_4$ & $T_{\mathrm{c}}$ \\
\hline
12.4\,K & 11.8\,K & 12.0\,K & 12.0\,K & 12.1\,K \\
\hline
\end{tabular}
\caption{Measured transition temperatures of the samples in Kelvin.}
\label{tab:tempdata}
\end{table}

\subsection*{Gating of the supercurrent}
\begin{figure}[]
    \centering
    \includegraphics[width=0.6\columnwidth]{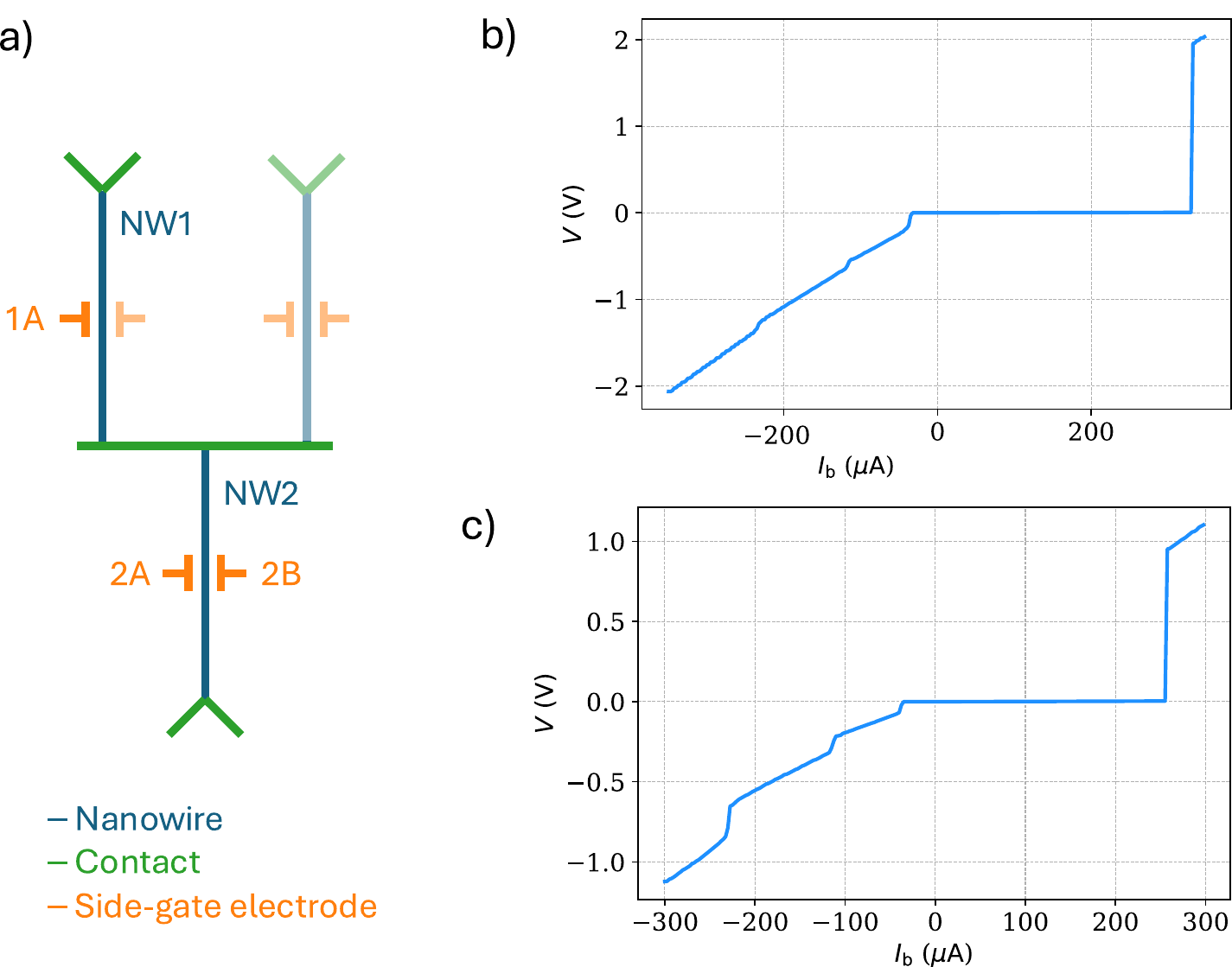}
    \captionsetup{labelformat=empty}
    \caption{SFig. 1. \textbf{Schematic and $V - I$ curves of the NWs used for the logic operations.} (a) Schematic of the device highlighting NW1 and NW2 and the side-gate electrodes. (b) $\mathrm{V-I}$ curve for NW1 at $V_{\mathrm{sg}}=0$. (c) $\mathrm{V-I}$ curve for NW2 at $V_{\mathrm{sg}}=0$.}
    \label{fig:sfig1}
\end{figure}

In our gating measurements we used only two nanowires, in this part we will refer to them as nanowire 1 (NW1) and nanowire 2 (NW2). NW1 is the left and NW2 is the lower nanowire from the network as depicted in SFig.~\ref{fig:sfig1}a). 
We first investigated the transport at $V_{\mathrm{sg}}=0$, for both nanowires. 
The $\mathrm{V-I}$ curves recorded for NW1 and NW2 at 4.2 K are shown in SFig.~\ref{fig:sfig1}b-c) respectively and the resistance values for the dissipative states for $I_{\mathrm{b}}>0$, as well as the maximal switching current values $I_{\mathrm{c1}}$ and $I_{\mathrm{c2}}$ are collected in Table \ref{tab:wiredata}.

\begin{table}[h!]
\centering
\begin{tabular}{|c|c|c|c|}
\hline
 & $I_{\mathrm{c}}\,(\mu\mathrm{A})$ & $I_{\mathrm{r}}\,(\mu\mathrm{A})$ & $R_{\mathrm{n}}\,(k\Omega)$ \\
\hline
NW1 & 332 & 35 & 5.86 \\
NW2 & 252 & 37 & 3.69 \\
\hline
\end{tabular}
\caption{Electrical characteristics of the nanowires.}
\label{tab:wiredata}
\end{table}

\begin{figure}[]
    \centering
    \includegraphics[width=0.75\columnwidth]{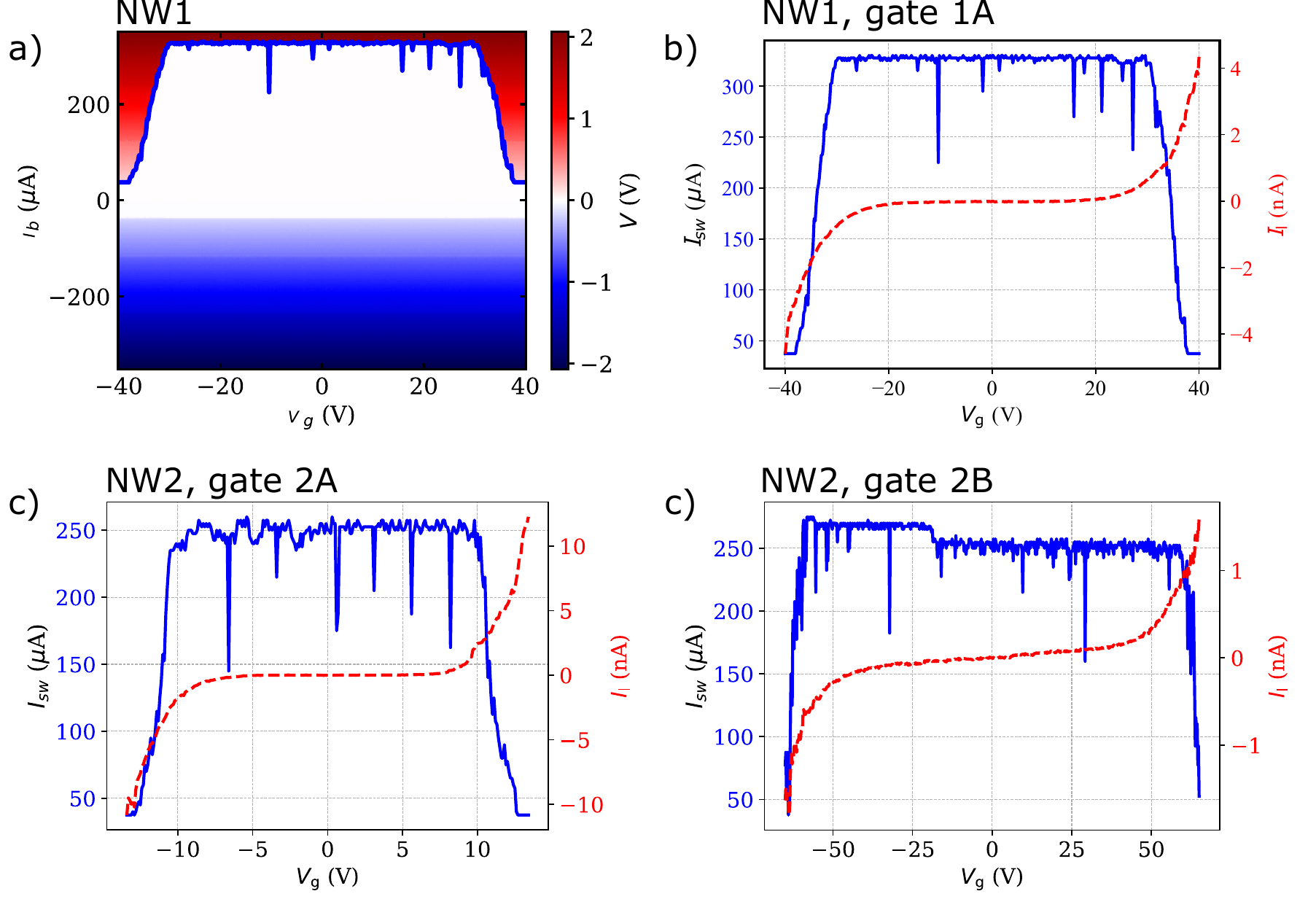}
    \captionsetup{labelformat=empty}
    \caption{SFig. 2. \textbf{Gating of the supercurrent} (a) Heat meap of the gating of NW1. (b) $I_{\mathrm{sw}}-V_{\mathrm{sg}}$ and $I_{\mathrm{l}}-V_{\mathrm{sg}}$ curves for NW1 and gate 1A. (c) $I_{\mathrm{sw}}-V_{\mathrm{sg}}$ and $I_{\mathrm{l}}-V_{\mathrm{sg}}$ curves for NW2 and gate 2A. (d) $I_{\mathrm{sw}}-V_{\mathrm{sg}}$ and $I_{\mathrm{l}}-V_{\mathrm{sg}}$ curves for NW2 and gate 2B.}
    \label{fig:sfig2}
\end{figure}

As a next step, we investigated the effect of the gate voltage applied to the side-gate electrodes.
In the gating experiments, we ramped the gate voltage from negative values to positive values, and recorded the $\mathrm{V-I}$ curve at each instant of the gate voltage set.
We show the heat map for NW1 in SFig.~\ref{fig:sfig2}a). 
The blue and red regions correspond to the dissipative states for negative and positive bias respectively, while the white part corresponds to the superconducting state.
For NW2 we show the maximal switching current as a function of the gate voltage applied to the side gate electrodes 2A and 2B, respectively, in SFig.~\ref{fig:sfig2}c-d), and we show the gating curve for gate 1A of NW1 in SFig.~\ref{fig:sfig2}a).

In order to perform logic operations, it is required that the maximal supercurrent of a nanowire changes significantly in a relatively small voltage window.
Ideally, a full suppression of the maximal supercurrent for $\approx 5\,$V would be optimal, so that two logical states corresponding to 0 and $\approx 5\,$V could be defined.
As the gate voltage required for the suppression of superconductivity in our devices is relatively large, we need an operation principle that can overcome this issue. 
Table \ref{tab:gating_data} shows the voltage values where the applied gate voltage stasrts to suppress the switching current and where is saturated, These values will be referred as $V_{\mathrm{sg,onset}}$ and $V_{\mathrm{sg,offset}}$. 
Table \ref{tab:gating_data} also shows the values of the maximal supercurrent at saturation, this value is denoted with $I_{\mathrm{s,sup}}$, the relatively large value is associated to the strong cooling power of the liquid helium.

\begin{table}[h!]
\centering
\renewcommand{\arraystretch}{1.2}
\begin{tabular}{|c|c|c|c|c|c|}
\hline
 & $V_{\mathrm{g,onset+}}\,(\mathrm{V})$ & $V_{\mathrm{g,offset+}}\,(\mathrm{V})$ & 
   $V_{\mathrm{g,onset-}}\,(\mathrm{V})$ & $V_{\mathrm{g,offset-}}\,(\mathrm{V})$ & 
   $I_{\mathrm{s,sup}}\,(\mu\mathrm{A})$ \\
\hline
NW1, 1A & 29.8 & 37.8 & $-30.0$ & $-37.8$ & 37.5 \\
NW2, 2A & 9.3 & 12.7 & $-9.4$ & $-12.9$ & 37.5 \\
NW2, 2B  & 58.6 & -- & $-59.6$ & -- & -- \\
\hline
\end{tabular}
\caption{Onset and offset gate voltages and the saturation of the maximal supercurrent.}
\label{tab:gating_data}
\end{table}

We show this values in Table \ref{tab:gating_data}, an additional $+/-$ indicates the polarity in the lower indicies of the respective quantities listed in the header of the table.
For NW2 gate electrode 2B we did not observe the saturation within the applied gate voltage range so the offset value of the gate voltage and the saturation value of the maximal supercurrent are not tabulated.

\begin{figure}[]
    \centering
    \includegraphics[width=0.4\columnwidth]{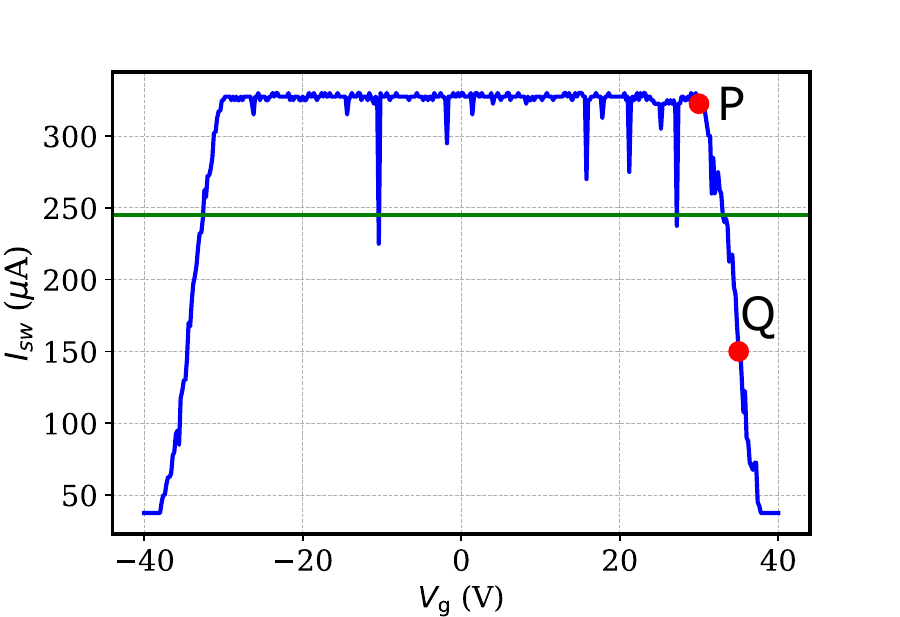}
    \captionsetup{labelformat=empty}
    \caption{SFig. 3. \textbf{Gating of the supercurrent and logic operation principle}  (b) $I_{\mathrm{sw}}-V_{\mathrm{sg}}$ for NW1 and gate 1A highlighting the points P and Q. Point P is in the vicinity of the onset of the action of the applied gate voltage, while point Q is already on the part of the gating curve, where the change in the maximal supercurrent is significant.}
    \label{fig:sfig3}
\end{figure}

Now it is clear that the gating characteristics is suitable for logic operations.
We demonstrate this in SFig.~\ref{fig:sfig3}, where we plot the maximal supercurrent as the function of the side gate voltage for NW1. 
First of all we set the bias current to $I_{\mathrm{b}}=245\,\mathrm{\mu A}$, as it is shown with a green solid line, and we define two points on the gating curve marked with P and Q respectively. 
Point P is in the vicinity of onset of the action of the gate on the maximal supercurrent and point Q corresponds to a gate voltage value which is greater by 5 V, as it is tabulated in Table \ref{tab:vg_is_data} together with the corresponding maximal supercurrent value.

\begin{table}[h!]
\centering
\begin{tabular}{|c|c|c|}
\hline
 & $V_{\mathrm{sg}}\,(\mathrm{V})$ & $I_{\mathrm{s}}$ \\
\hline
P & 30 & 322.5 \\
Q & 35 & 150 \\
\hline
\end{tabular}
\caption{Gate voltages and corresponding supercurrents at points Q and P.}
\label{tab:vg_is_data}
\end{table}

If we initially prepare the system at point P and then apply an additional 5\,V gate voltage on the gate electrode we can switch to point Q.
If we want to switch the device back to the superconducting state, then we need to reset the value of the voltage applied to the side-gate electrode. 
Due to the relatively large hysteresis of the device, at this point of the operation we also need to set the bias current below the retrapping current (or simply to 0) for a short amount of time, as we already discussed in the corresponding section of the main text (2. B. Logic operation principle).

\section*{\large Logic gate measurements}
In this section we present a more detailed description of the operation of the logic gates including the bias current values, and the working points on the $I_{\mathrm{sw}}-V_{\mathrm{sg}}$ curves.
\subsection*{AND gate}
In this measurement NW1 was gated with signal inputA on gate electrode 1A and NW2 was gated with signal inputB on gate electrode 2A. 
The chosen working points and the bias current $I_{\mathrm{b}}$ for NW1 and NW2 are tabulated in Table \ref{tab:and_table}, $V_{\mathrm{g,wpA}}$ and $V_{\mathrm{g,wpB}}$ denote the points on the $I_{\mathrm{sw}}-V_{\mathrm{sg}}$ curve close to the vicinity of the onset of the gating effect, while $V_{\mathrm{g,pulseA}}$ and $V_{\mathrm{g,pulseB}}$ represent points on the respective $I_{\mathrm{sw}}-V_{\mathrm{sg}}$ curves of the nanowires where the maximal supercurrent is already suppressed.
The operation utilizes a logical signal with a relatively low amplitude of $5\,\mathrm{V}$.

\begin{table}[h!]
\centering
\begin{tabular}{|c|c|c|c|c|c|c|}
\hline
$I_{\mathrm{b}}\,(\mu\mathrm{A})$ & $V_{\mathrm{g,wpA}}\,(\mathrm{V})$ & $V_{\mathrm{g,pulseA}}\,(\mathrm{V})$ & $V_{\mathrm{g,wpB}}\,(\mathrm{V})$ & $V_{\mathrm{g,pulseB}}\,(\mathrm{V})$ & Output (V) \\
\hline
175 & 29.85 & 34.85 & 9.95 & 14.95 & 0.47 \\
\hline
\end{tabular}
\caption{Gate voltages, bias current, and output voltage for the AND gate.}
\label{tab:and_table}
\end{table}

\subsection*{OR gate}
Here we present an alternative realization of the OR circuit with the same circuit that we used for the AND gate, but the values of the bias current and the applied gate voltage to the side gate electrode has to be changed, as we tabulated in Table \ref{tab:or2_table}.
At this point it is worth to remember that we use 5 V signal all the time, and changing the working point changes the maximal voltage on the gate, which influences the critical current of the wire. In the region defined by the working points, the critical current is sensitive to the change of the gate voltage. We do not suppress the critical current totally in both cases, and this is why we get OR functionality instead of AND with varying the working point.

\begin{figure}[H]
    \centering
    \includegraphics[width=0.9\columnwidth]{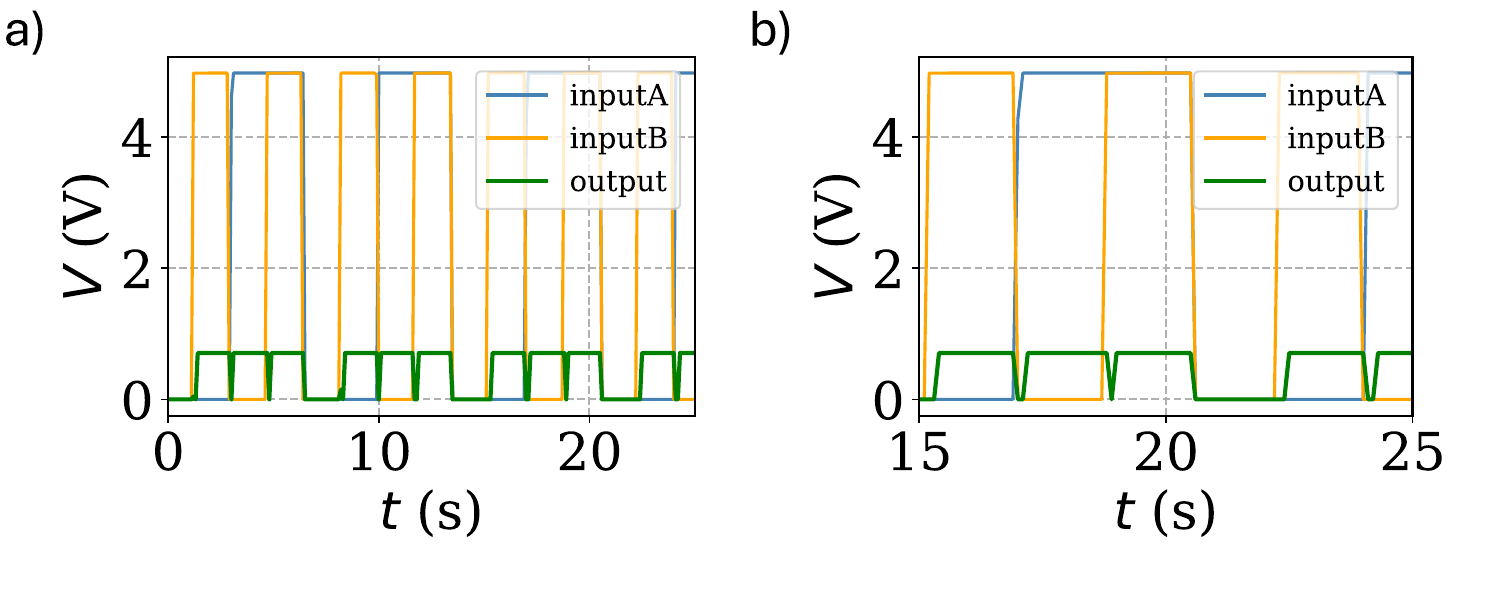}
    \captionsetup{labelformat=empty}
    \caption{SFig. 4. \textbf{Realization of the OR gate with the same setup used for the AND gate} (a) The inputs and the output as the funciton of time. (b) Two cycles of the operation of the OR gate.}
    \label{fig:sfig4}
\end{figure}

With the application of these parameters we can achieve such operation of our device that the truth table of the OR gate is reproduced in terms of the output voltage as it is demonstrated in SFig.~\ref{fig:sfig4}.

In the case when both inputs are zero (with respect to the chosen working point taken as offset) both nanowires are in the superconducting state and the output will be zero. 
In the case, when any of the inputs is large (the extra $5\,\mathrm{V}$ logical signal is applied) one of the wires will switch to the normal state and the truth table of the OR gate is reproduced in terms of the measured output voltage.

\begin{table}[h!]
\centering
\begin{tabular}{|c|c|c|c|c|c|}
\hline
$I_{\mathrm{b}}\,(\mu\mathrm{A})$ & $V_{\mathrm{g,wpA}}\,(\mathrm{V})$ & $V_{\mathrm{g,pulseA}}\,(\mathrm{V})$ & $V_{\mathrm{g,wpB}}\,(\mathrm{V})$ & $V_{\mathrm{g,pulseB}}\,(\mathrm{V})$ & Output (V) \\
\hline
232.5 & 34.12 & 39.10 & 9.92 & 14.92 & 0.47 \\
\hline
\end{tabular}
\caption{Gate voltages, bias current, and output voltage for the OR gate.}
\label{tab:or2_table}
\end{table}

For the second realization of the OR gate, presented in the main text the carefully chosen working points for NW2 and the bias current $I_{\mathrm{b}}$ are tabulated in Table \ref{tab:or1_table}, $V_{\mathrm{g,wpA}}$ and $V_{\mathrm{g,wpB}}$ denote the points on the $I_{\mathrm{sw}}-V_{\mathrm{sg}}$ curves close to the vicinity of the onset of the gating effect for the two sides gates of NW2, while $V_{\mathrm{g,pulseA}}$ and $V_{\mathrm{g,pulseB}}$ represent points on the respective $I_{\mathrm{sw}}-V_{\mathrm{sg}}$ curves of NW2 where the maximal supercurrent is already suppressed for the different side-gate electrodes.

\begin{table}[h!]
\centering
\begin{tabular}{|c|c|c|c|c|c|}
\hline
$I_{\mathrm{b}}\,(\mu\mathrm{A})$ & $V_{\mathrm{g,wpA}}\,(\mathrm{V})$ & $V_{\mathrm{g,pulseA}}\,(\mathrm{V})$ & $V_{\mathrm{g,wpB}}\,(\mathrm{V})$ & $V_{\mathrm{g,pulseB}}\,(\mathrm{V})$ & Output(V) \\
\hline
210 & 8.0 & 13.0 & 58.0 & 63.0 & 0.86 \\
\hline
\end{tabular}
\caption{Gate voltages, bias current, and output voltage for the OR gate.}
\label{tab:or1_table}
\end{table}

\subsection*{NOT gate}

In this measurement NW1 was gated with signal inputA on gate electrode 1A and the output was measured on NW2 and an additional ohmic resistance of $1\,\mathrm{k\Omega}$ was put in parallel with the nanowires.
The chosen working point and the bias current $I_{\mathrm{b}}$ for NW1 are tabulated in Table \ref{tab:not_table}, $V_{\mathrm{g,wp}}$ denote a point on the $I_{\mathrm{sw}}-V_{\mathrm{sg}}$ curve close to the vicinity of the onset of the gating effect, while $V_{\mathrm{g,pulse}}$ represents a point on the $I_{\mathrm{sw}}-V_{\mathrm{sg}}$ curve of the nanowire where the maximal supercurrent is already suppressed.

\begin{table}[H]
\centering
\begin{tabular}{|c|c|c|c|c|}
\hline
$I_{\mathrm{b}}\,(\mu\mathrm{A})$ & $V_{\mathrm{g,wpA}}\,(\mathrm{V})$ & $V_{\mathrm{g,pulseA}}\,(\mathrm{V})$ & Output (V) \\
\hline
260 & 30.0 & 35.0 & 0.16 \\
\hline
\end{tabular}
\caption{Gate voltages, bias current, and output voltage for the NOT gate.}
\label{tab:not_table}
\end{table}

\subsection*{COPY gate}
In this measurement NW1 was gated with signal inputA on gate electrode 1A.
The chosen working point and the bias current $I_{\mathrm{b}}$ for NW1 are tabulated in Table \ref{tab:copy_table}, $V_{\mathrm{g,wp}}$ denote a point on the $I_{\mathrm{sw}}-V_{\mathrm{sg}}$ curve close to the vicinity of the onset of the gating effect, while $V_{\mathrm{g,pulse}}$ represents a point on the $I_{\mathrm{sw}}-V_{\mathrm{sg}}$ curve of the nanowire where the maximal supercurrent is already suppressed.

\begin{table}[h!]
\centering
\begin{tabular}{|c|c|c|c|}
\hline
$I_{\mathrm{b}}\,(\mu\mathrm{A})$ & $V_{\mathrm{g,wp}}\,(\mathrm{V})$ & $V_{\mathrm{g,pulse}}\,(\mathrm{V})$ & Output(V) \\
\hline
200 & 31.84 & 33.53 & 1.72 \\
\hline
\end{tabular}
\caption{Gate voltages, bias current, and output voltage for the COPY gate.}
\label{tab:copy_table}
\end{table}
\newpage
\section*{\huge Universality of the circuit}
In this section, we present the schematic of other classical logic gates and computational circuits and discuss their operation briefly.
We also present calculated input-output curves for the different circuits. The resistance of the two parallel (upper) nanowires were chosen to be $R_2 = \SI{15}{k\ohm}$, the resistance of the third (lower) nanowire was $R_1 = \SI{4}{k\ohm}$, the external resistor was $r=\SI{200}{\ohm}$, and the critical current of the nanowires is $I_0=\SI{43}{\mu\ampere}$.

\begin{figure}[H]
    \centering
    \includegraphics[width=0.75\columnwidth]{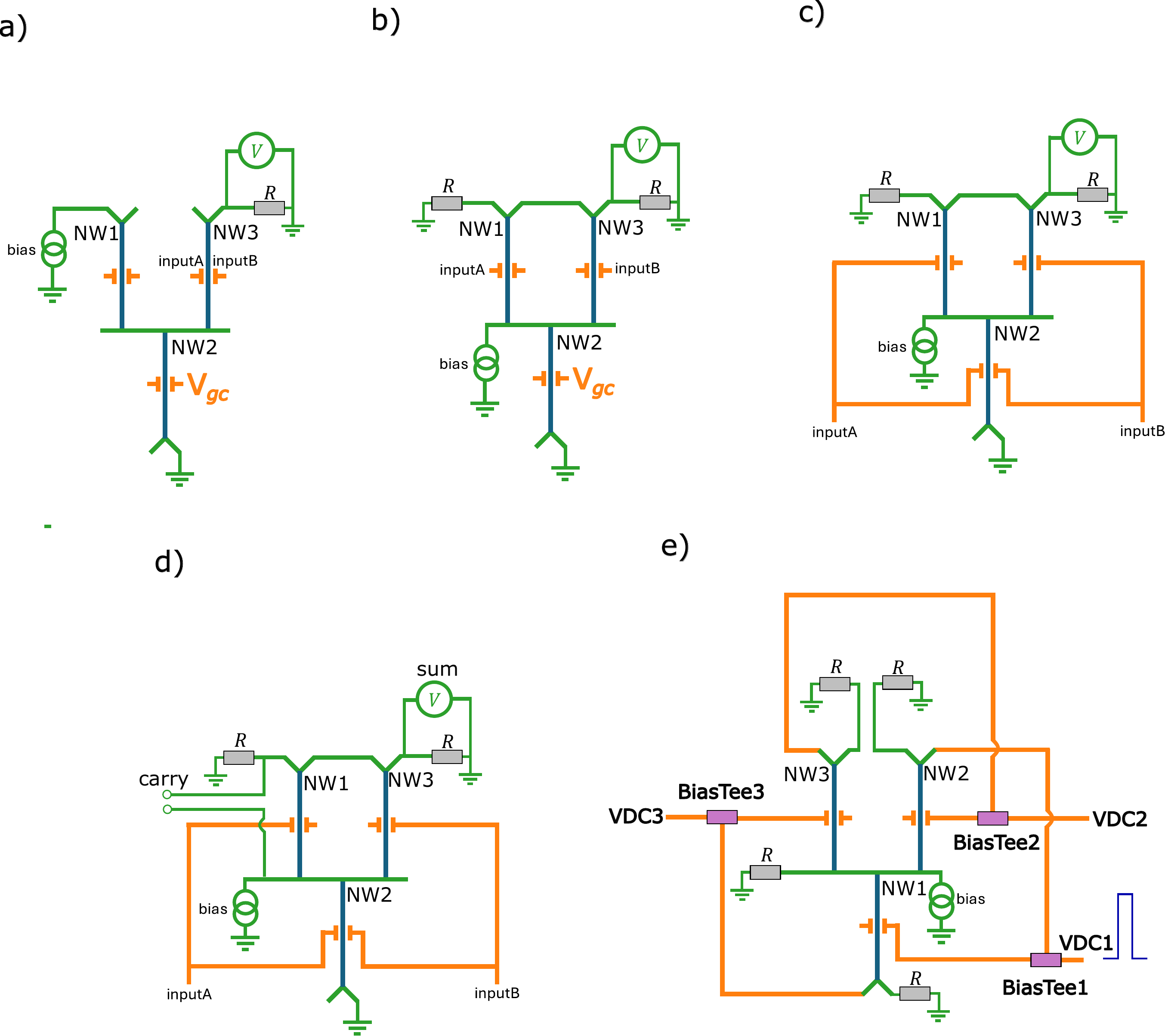}
    \captionsetup{labelformat=empty}
    \caption{SFig. 5. \textbf{Realisation of the NOR, NAND, XOR logic gates, the HALF ADDER combinational circuit and the RING OSCILLATOR circuit.}Panels a)-e) show the circuit configurations for the classical logic gates and (combinational) circuits not presented in the main text.}
    \label{fig:sfig5}
\end{figure}

\subsection*{NOR}
The circuit configuration for the NOR gate is shown in SFig.~\ref{fig:sfig5}a).
In case of the NOR gate, NW2 is kept normal by applied gate voltage $V_{\text{gc}}$, the logic signals inputA and inputB are applied on the two side-gate electrodes on NW3, and finally the system is current biased through NW1 as shown in the figure. The output will be low only if both the second and third arm is switched to the N state. In the case when one or both of the nanowires are in the SC state they will shunt each other and the output voltage will be high. The input-output of the circuit as the function of time is shown in SFig.~\ref{fig:sfig6}a).

\subsection*{NAND}
The circuit configuration for the NAND gate is shown in SFig.~\ref{fig:sfig5}b).
In case of the NAND gate, NW2 is kept normal by the applied gate voltage $V_{\text{gc}}$, the logic signals inputA and inputB are applied on one of the side-gate electrodes of NW1 and NW2, and finally the system is current biased through commmon line as shown in the figure. The output will be low only if all the nanowires are switched to the N state. In the case when any of the two nanowires gated with logic signals are in the SC state, they will shunt each other, and the output voltage will be high. The input-output of the circuit as the function of time is shown in SFig.~\ref{fig:sfig6}b).

\subsection*{XOR}
The circuit configuration for the XOR gate is shown in SFig.~\ref{fig:sfig5}c).
The principle of operation for the XOR gate is the same as for the NAND gate, but in this case the NW2 is controlled by both inputA and inputB as shown in the figure.  If both inputs are low, then the third nanowire will be in the SC state and the output will be low, however if both inputs are high, then all the nanowires are switched to the N state and the current will be distributed in the three arms. In all the other cases the first and second arms will shunt the first arm, and the output will be high due to the increased current. The input-output of the circuit as the function of time is shown in SFig.~\ref{fig:sfig6}c).

\subsection*{HALF ADDER}
\begin{figure}[]
    \centering
    \includegraphics[width=1.\columnwidth]{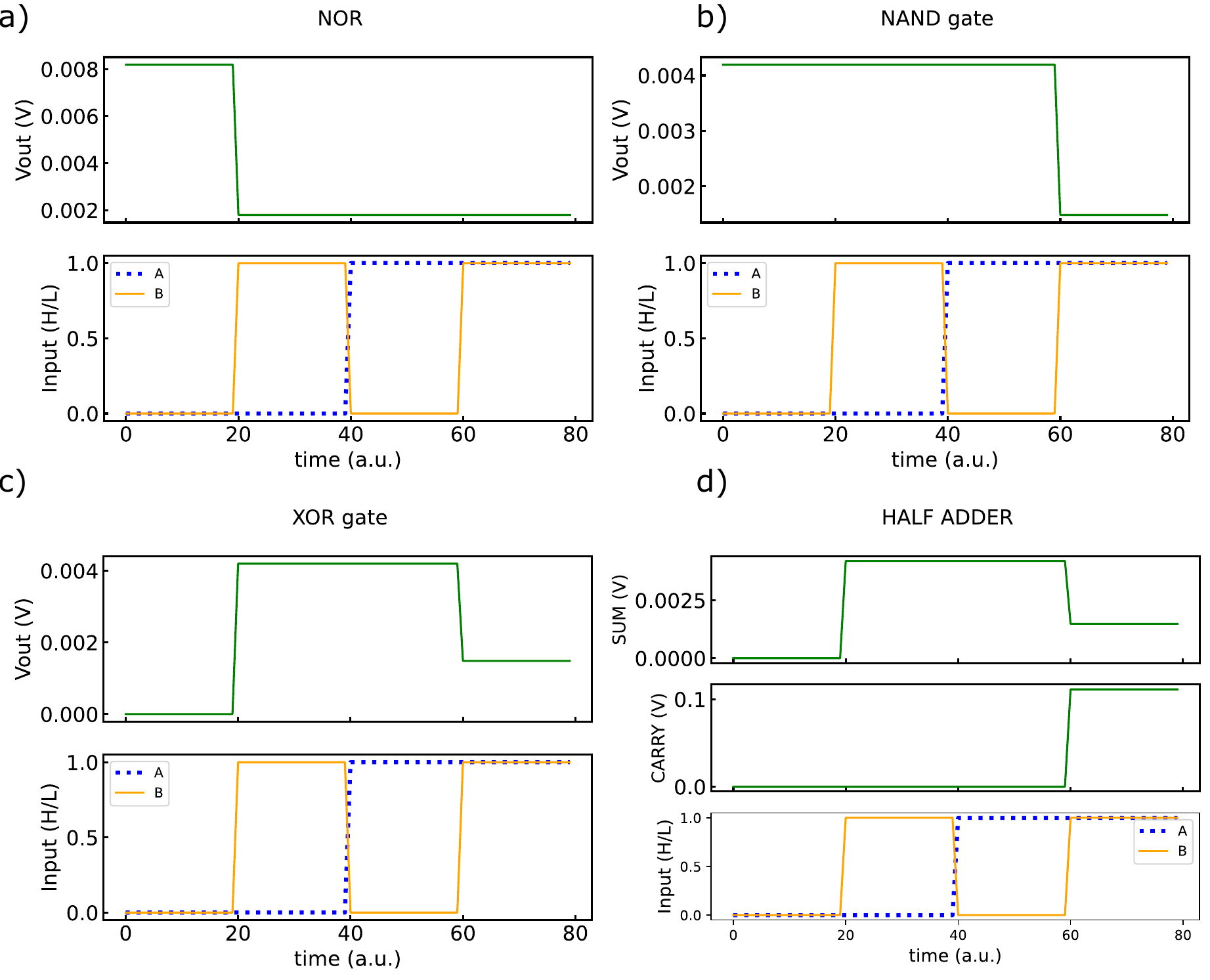}
    \captionsetup{labelformat=empty}
    \caption{SFig. 6. \textbf{Predicted output of the NOR, NAND, XOR logic gates and the HALF ADDER combinational circuit.}Panels a)-d) show the input-output as the function of time for the classical logic gates and (combinational) circuits not presented in the main text.}
    \label{fig:sfig6}
\end{figure}

The circuit configuration for the HALF ADDER combinational circuit is shown in SFig.~\ref{fig:sfig5}d).
The idea is the same as for the XOR gate, the only difference is that we measure the CARRY line on NW1, as it is shown in the figure. The CARRY output indicates that the addition result is too high for one bit and can be passed further to the next higher bit position.
The SUM is measured in the same way, as in case of the XOR gate, and the theoretically predicted logical states of the output voltage are also the same. The input-output of the circuit as the function of time is shown in SFig.~\ref{fig:sfig6}d).

\subsection*{RING OSCILLATOR}
The circuit configuration for the RING OSCILLATOR circuit is shown in SFig.~\ref{fig:sfig5}e), take care, that the NWs are relabeled on thsi figure.
In this circuit configuration, additional Bias Tees are used.
Since the ring oscillator can be built with 2n + 1 NOT gates, the Y-junction configuration can satisfy this condition with n=1. 
If the three nanowires are connected in series with three external resistors and in parallel with a fourth one as indicated in the figure, three NOT gates can be realised.
In this circuit configuration, the "output signal" from each gate is connected to the input of the next gate through a bias tee BiasTee2, BiasTee3 and BiasTee1 respectively.
This connection results in a "coupling", as the "output voltage" on NW3 is connected to the bias tee BiasTee2 corresponding to NW2.
The "output voltage" on NW2 is connected to BiasTee1 corresponding to NW1.
The "output voltage" on NW1 is connected to BiasTee3 corresponding to NW3, and with this the "loop" is closed.
The bias tees BiasTee1, BiasTee2 and BiasTee3 are used to keep the potential of the gate higher than that of the circuit by the DC bias voltage VDCi where i=1,2,3 respectively.
If a signal on top of the DC signal for e.g. the gate electrode is applied, as shown in the figure, then the measured output will oscillate between two voltage levels with some time constant.

\end{document}